# Label-free three-dimensional (3D) structural imaging in live cells using intrinsic optical refractive index

**Authors:** Chen Liu[1,2,6], Michael Malek[2], Ivan K. H. Poon[3], Lanzhou Jiang[3], Colin J. R. Sheppard[4], Ann Roberts[5], Harry Quiney[5], Douguo Zhang[6], Xiaocong Yuan[1*], Jiao Lin[1,5,7], Christian Depeursinge[8,9*], Pierre Marquet[10,11,12*], Shan Shan Kou[2,5,8]*

**Affiliations:**

[1]Nanophotonics Research Centre, Shenzhen University & Key Laboratory of Optoelectronic Devices and Systems of Ministry of Education and Guangdong Province, College of Optoelectronic Engineering, Shenzhen University, Shenzhen 518060, China.

[2]Department of Chemistry and Physics, La Trobe Institute for Molecular Sciences (LIMS), La Trobe University, Melbourne, Victoria 3086, Australia.

[3]Department of Biochemistry and Genetics, La Trobe Institute for Molecular Science (LIMS), La Trobe University, Melbourne, Victoria 3086, Australia.

[4]Istituto Italiano di Tecnologia, Genova 16163, Italy.

[5]School of Physics, University of Melbourne, VIC 3010 Australia.

[6]Department of Optics and Optical Engineering, Anhui Key Laboratory of Optoelectronic Science and Technology, University of Science and Technology of China, Hefei, Anhui 230026, China.

[7]School of Engineering, RMIT University, Melbourne, VIC 3001, Australia.

[8]Microvision and Microdiagnostic Group (SCI STI CHD), Ecole Polytechnique Fédérale de Lausanne (EPFL), 1015 Lausanne, Switzerland.

[9]Laboratory for Cellular Imaging and Energetics, Biological and Environmental Sciences and Engineering Division, King Abdullah University of Science and Technology (KAUST), Thuwal 23955-6900, Kingdom of Saudi Arabia.

[10]Joint International Research Unit in Neurodevelopment and Child Psychiatry, CHUV, Département de Psychiatrie, Lausanne, Switzerland, Université Laval ,Québec, Québec, Canada.

[11]Institut universitaire en santé mentale de Québec, Québec, Québec, Canada.

[12]Centre d'optique, photonique et laser, Université Laval, Québec, Québec, Canada, Department of Psychiatry & Neuroscience, Université Laval, Québec, Québec, Canada.

*Correspondence to: xcyuan@szu.edu.cn, christian.depeursinge@epfl.ch, Pierre.Marquet@neuro.ulaval.ca & s.kou@latrobe.edu.au

**Abstract:** Here we report a method for visualization of volumetric structural information of live biological samples with no exogenous contrast agents. The process is made possible through a technique that involves generation, synthesis and analysis of three-dimensional (3D) Fourier components of light diffracted by the sample. This leads to the direct recovery of quantitative cellular morphology with no iterative procedures for reduced computational complexity.

Combing with the fact that the technique is easily adaptive to any imaging platform and requires minimum sample preparation, our proposed method is particularly promising for observing fast, volumetric and dynamic events previously only accessible through staining methods.

**Main Text:** Imaging and visualization of the three-dimensional (3D) structural information in cellular and sub-cellular environments will profoundly influence the way we perceive the underlying mechanisms of the complex biological world. At present, 3D imaging techniques usually involve exogenous fluorescent dyes. A problem with using dyes, however, is that it is sometimes undesirable to stain cells for certain important biomedical research topics, including stem cell and fertility studies, and the process can become difficult when multiple fluorophores are involved. On the other hand, there has been considerable recent interest in developing high-resolution, 3D, label-free imaging techniques. Indeed, while fluorescent dye has the capacity to target specific cellular components, with a high degree of reliability, rapidly obtaining high-resolution volumetric data with intrinsic contrast is particularly desirable in providing complementary information about the background biological context (*1*). For label-free techniques, the image contrast is intrinsic, arising from spatial variations in the optical refractive index (RI denoted as *n*) of the object. Although label-free imaging techniques span various electromagnetic (EM) wavelengths including infrared (IR) and/or near-infrared (NIR), excitation using non-linear optics (*2*) and x-rays (*3,4*) as well as electrons (*5*), visible light (VIS) is more readily adaptable to existing biological laboratory infrastructure and is relatively non-destructive to biological samples. In this case, however, the effects of light diffraction need to be taken into account to obtain a more accurate representation of the propagation of light through such objects.

Here, we present a new approach to optical tomographic imaging where unique quantitative spatial mapping of subcellular structures is demonstrated. This enables localized functional interpretations in relative biomedical studies, hence bridging the significant gap in traditional imaging between function and structure. The proposed modality provides structural information in full 3D, with no iterative procedures nor any staining.

Instead of the classical "plane-by-plane" (2D) based calculation for 3D reconstruction of a light field, we mapped the 2D frequencies to a spherical shell in the 3D Fourier domain which represents the entire volumetric diffracted field (Fig. 1). A homogeneous Green's function containing essentially only the propagating waves (ignoring evanescent components) is considered as the Helmholtz operator (*6*), which is valid under most imaging conditions for biological cells. We have shown that it transforms to a spherical shell in the Fourier domain for monochromatic illumination (*7*). Thus, the 3D scattering potential of the object from one particular angular direction is analytically linked to the 3D angular spectrum by a thin spherical shell in the Fourier domain. This removes the need for any iterative procedures, and greatly simplifies the entire tomographic reconstruction procedure. We have demonstrated its feasibility with quantitative reconstructions of live cells using a simple holographic setup with 30 scanning angles at a 20x magnification (N.A. = 0.45). The reconstructed 3D structural information is also encoded with pseudo-colours which represent different regions of similar refractive index (RI) values. Typically a cell is a complex 3D object consisting of many subcellular components and organelles. It has been discovered that the RI values vary across these minute biological structures (Fig. S1). The value of the RI quantifies the speed of light in media which is related to the regional concentration of protein (which has a higher value of the refractive index). Encoding similar refractive indices is, therefore, an intuitive way to map out regions sharing similar biophysical content – a useful perspective that could greatly complement fluorescence-based imaging. Many of the RI values at the sub-cellular level remain unknown (*8*). This report could open doors to unprecedented discoveries leading to label-free functional imaging inside the cell.

In the proposed cellular tomography, a localized refractive index profile $n(\boldsymbol{r})$ is calculated from its direct relationship with the scattering potential as $F(\boldsymbol{r}) = k^2[n^2(\boldsymbol{r}) - 1]/4\pi$, where $\boldsymbol{r}$ is the 3D coordinate (*9*). This is valid under the condition of weak absorption, as in the case of visible light illuminating an unstained biological sample. Obtaining the scattering potential is an inverse procedure where parts of its 3D Fourier transform $\tilde{F}(\boldsymbol{K})$ are firstly synthesized in the conjugate K-space from various angular conditions (Fig. 2). When the 3D spatial information of the object, carried in the form of Fourier spectra, is mapped correctly into K-space, an inverse 3D Fourier transform restores the object's scattering potential, $F(\boldsymbol{r})$. The key to the problem is how to synthesize the complex spectrum of the scattering potential, $\tilde{F}(\boldsymbol{K})$, correctly into K-space. In the existing theoretical framework of ODT, the analytical object space is broken into two $z^+$ and $z^-$ half-spaces (Fig. S2). The 3D scattering potential of a scatterer, therefore, may be found using the 2D Fourier components of the scattered field, provided that the scattering distance $z^+$ is known (*9*). However, in a microscopic set-up with an amorphous sample such as a cell, the true scattering distance of an object is often difficult to measure or define precisely, making it hard to implement this method directly for cellular imaging (*10*). We take a different approach: the 3D scattering potential is synthesized with 3D Fourier components of the scattered field, meaning that the analytical object space is not broken into two half spaces but is rather considered as a 3D entity directly. Because the entire theoretical framework is derived in 3D instead of considering the scattering spaces into two hemispheres, no iterative plane-by-plane "back-propagation" is needed (since the effects of diffraction are *inherently* accounted for using the spherical support in K-space). The spectrum of the scattering potential $\tilde{F}(\boldsymbol{K})$ of the object is constructed in 3D according to the angular geometry (*11*) (Fig. 2), and inversely retrieved by a single Fourier transform. The detailed derivation can be found in the supplementary text. A difference between our method and that of computed tomography (CT) (*12*) is that, in our approach, the information from one monochromatic image must be transcribed to the conjugate K-space under the spatial support of a thin spherical cap, instead of a plane, due to the effect of light diffraction (Fig. S3). In addition, our method has enhanced accuracy due to the fact that there is no need to perform any phase unwrapping that is commonly found in 2D-based reconstruction methods.

The principles described previously apply universally to any type of tomography in the case of weak scattering, meaning that it could be applied equally well to the cases of x-ray or electron waves when scattering is weak but not negligible. To demonstrate the method's robustness in cellular imaging, we constructed a bench-top cellular tomography system with varied illumination scan patterns. The set-up is based on a Mach-Zehnder interferometer, with the addition of a scanning mechanism (Fig. S4A). Different types of human cells are prepared according to standard culturing protocols (Supplementary Methods) and 3D mappings of the RI distributions are recovered using a simple linear scan pattern. We imaged firstly the primary human buccal epithelial cell where regions of different refractive index are colour-coded (Fig. 3A). The reconstruction is completely three-dimensional and we are able to generate a rotating view of the sample object (Supplementary Movie 1). The shape of the nucleus can be clearly delineated. Cytoplasmic components are divided into two layers through segmentation of the values of the refractive indices. This reveals interesting morphological behavior of cytoskeletal remodeling that occurs at the edge of a cell, which is usually unseen when cells are unlabeled under conventional microscopes. We also explored more samples such as neurons to exhibit the important capability of the proposed technique – to unveil functionality and morphologies with no labeling. This is clearly reflected in the reconstructed neuron sample (Fig. 3C), which reveals spatial distributions of clustering in organelles with similar biomass (Supplementary Movie 2). The higher RI value (represented with red color in Fig. 3C) shows a hollow shape which cannot be revealed in traditional 2D phase image result (Fig. 3D), giving a good example of how this method isolates the variation of thickness and RI value of the sample.

To further investigate the potential of the proposed method in practical biological study, the tomography system was combined with epi-fluorescence microscopy components (Fig. S4B) to obtain the distributions of RI and fluorescence intensities (a measure of nuclear contents here) in the same measured area to study the distribution of nuclear contents during the cell death process. Human Jurkat T cells were prepared according to mammalian cell culturing protocols, and stained with Hoechst 33342 dye to monitor the localization of nuclear contents before imaging or induction of apoptosis (a form of programmed cell death) (Supplementary Methods). 3D mappings of the RI distributions of Jurkat T cells were recovered (Fig. 4A) and the distributions of wide-field fluorescence intensities were obtained simultaneously (Fig. 4C). The 3D RI image shows more clearly than the fluorescent counterpart that the nuclear contents condensed and present a hollow shape (a well described hallmark of apoptosis (*13*)) (Supplementary Movie 3). The phenomenon of chromatin condensation can be easily observed through a 2D slice of RI distributions along the z-axis (Fig. 4B). Our method is compatible with the conventional fluorescence microscope method. Furthermore, when combined, we can enhance the fluorescence images by revealing unseen morphological changes that could stimulate interpretation of the underlying biological mechanism. We can access the interior of the cell and show its detailed interior structures (Fig. 4A). Although such a result could be achieved with Z-stack imaging of staining, our method shows even more insight as it can provide the entire spatial information of the cell and its subcellular structures potentially in real time, while fluorescence imaging needs the introduction of multi-stain and an extended period of exposure time, which could damage the cell. We obtained further comparisons between RI values and fluorescence intensities for Jurkat T cells under different conditions. Similar results were achieved and shown in Supplementary Fig. 5.

To demonstrate the quantitative nature of the method, we simulated tomographic images of a 3D scattering object using Monte Carlo principles and reconstructed the phantom using the above principles (Supplementary Text and Supplementary Fig. S6). Our testing object is a 3D semi-spherical shape in a 3D environment resembling typical cellular immersion under physiological conditions. Volumetric diffraction within the computational boundary is derived from a numerical implementation of Born's diffraction integral. The correct RI values and the shape of the phantom were recovered validating our method. From the simulation, we also observed an important phenomenon: only a handful of scan angles are needed for a complete reconstruction. Reducing the N.A. affects the resolution of reconstruction, but not the recovered RI values. This makes the proposed technique suitable even when only a limited amount of data is acquired potentially enabling rapid evaluations in live cell studies.

In conclusion, we have demonstrated a cellular imaging method that elucidates entire 3D sub-cellular structures with no staining nor specialized preparation. The system works with both adherent and non-adherent cells, which can be very promising for stem cells and organelles that prefer suspension environment. Furthermore, the algorithm is independent of the details of the experimental platform as long as angular information is available.  These properties indicate that the proposed imaging modality a powerful tool that could be readily adapted to any biological laboratory setting.

**Acknowledgments:**

This work was largely supported by the start-up grant of S.S.K. from La Trobe Institute of Molecular Sciences (LIMS). S.S.K. acknowledges the Discovery Early Career Researcher Award (DECRA) funded by the Australian Research and the Partnership Grant from La Trobe Research Focus Area (RFA) of Understanding Diseases.

Partial support of the project also came from the National Natural Science Foundation of China under Grant Nos.  61427819; Science and Technology Innovation Commission of Shenzhen under grant Nos. KQCS2015032416183980. X.-C. Yuan acknowledges the support given by the leading talents of Guangdong province program No. 00201505.

Contributions

SSK conceived the idea and designed the principles of the 3D microscope. CL and SSK built the instrumentation, performed the experiment, acquired and analyzed the data. MM, IP, LJ assisted with the experiments and data interpretation. CS, JL, CD and PM clarified the underlying physics and helped formulated the theoretical model.  SSK and CL drafted the manuscript with inputs from all co-authors. SSK, PM, CD, AR, HQ, DZ, XY supervised the experiments at respective stages and revised the manuscript critically. SSK directed the project. All authors discussed the results.

Competing interests

The techniques described in this manuscript are associated with an Australian provisional patent.

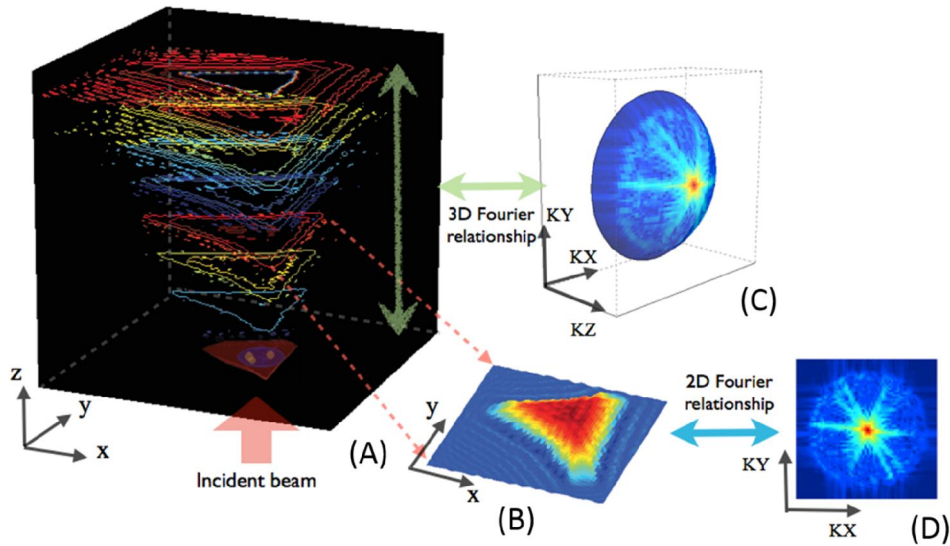

**Fig. 1. The principle of volumetric diffraction using a single 3D Fourier entity. (A) 3D light field of a cell, shown as color-coded planes of phase information. (B) 2D light field of a cell represented by the phase information. (C) 3D spectrum of the light field from a cell. (D) 2D spectrum of the light field from a cell.** In (A), the 3D volumetric diffracted field from a cell under a single incident angle is illustrated in a "slice-by-slice" manner, with different color codes for each plane. Each pseudo-color coded plane is an indicative line representation of the phase information, where one example plane is reconstructed in (B). In our technique, this entire 3D information can be retrieved from a single 3D Fourier transform, as all of the volumetric field information is synthesized onto the thin spherical support of a cap in K-space represented in (C). The "cap" is "scooped out" from a stack of 2D spectra. The N.A. of the cap is determined by the illumination angle, and the thickness of the cap related to the illumination wavelength. This technique is very different from the classical reconstruction of a light field using the 2D Angular Spectrum method where only planar information (B) can be recovered from the 2D spectrum of the object field (D).

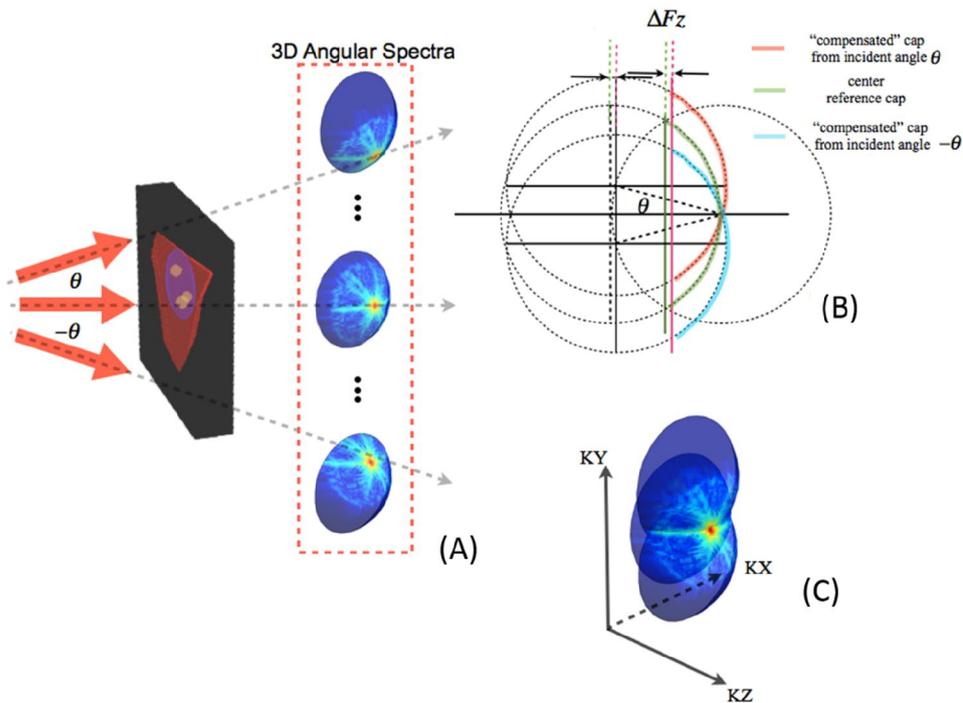

Fig. 2. Synthesis of the complex spectra of the scattering potential of a cell (A) A cell is subject to a multi-angle plane wave illumination with a maximum subtended half-angle as $\theta$, hence the numerical aperture of the system (N.A) is $\sin(\theta)$ (B) 2D side view of the angular compensation method introduced during the synthesis of the 3D Angular Spectrum (only three angles shown) (C) 3D view of a "synthesized" Angular Spectrum (with three example angles only). When a biological sample (such as a cell) is illuminated by a multi-angle beam, each illumination angle is considered separately and an individual cap of sphere is formed in the Fourier domain as the 3D Angular Spectrum. In (A), only the center illumination and two maximum subtended angles are shown. The respective 3D Angular Spectra are then translated and shifted accordingly, as shown in (B) from a side viewpoint, so that angular information introduced in the incident beams are compensated. $\Delta Fz$ is the amount of shift in axial direction for the incident angle of $\theta$. Synthesized Angular Spectrum for three example angles is shown in (C), where all the 0th order components coincide at a point.

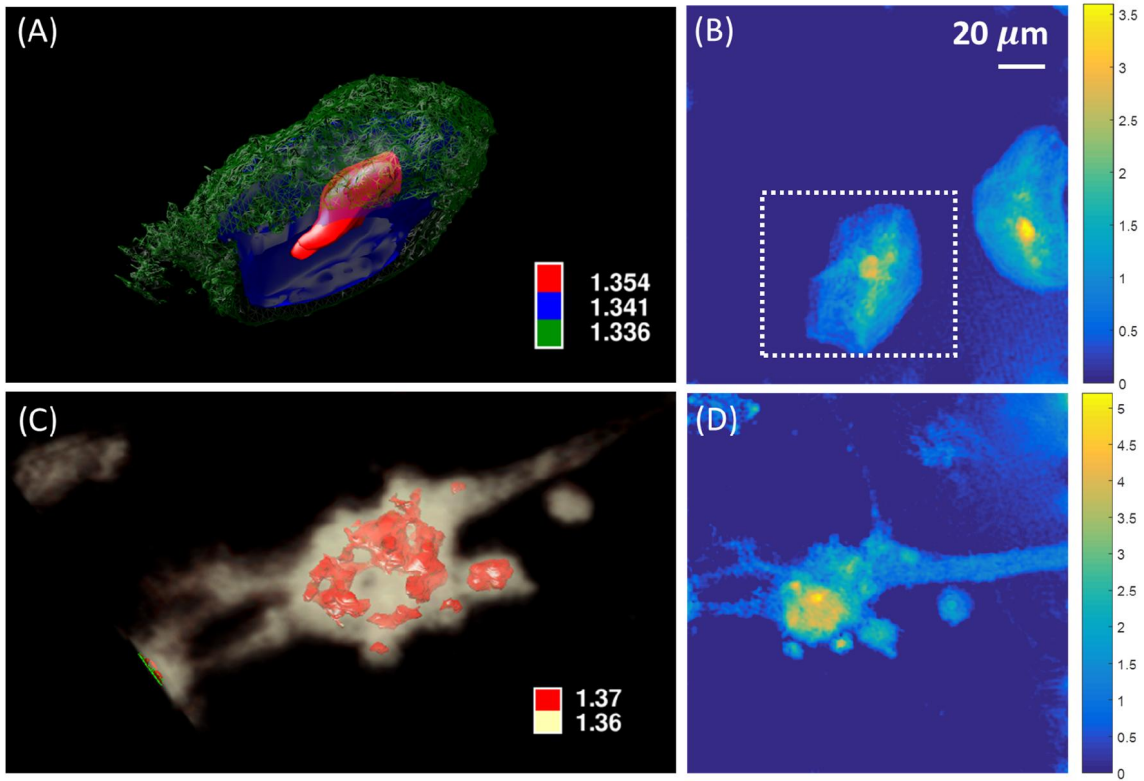

Fig. 3. Results stain-free (A) cheek cell and (C) neuron with internal structures delineated from groupings of refractive index values. Phase distribution of the same (B) cheek cell (highlighted with dash box) and (D) neuron (unit radian).

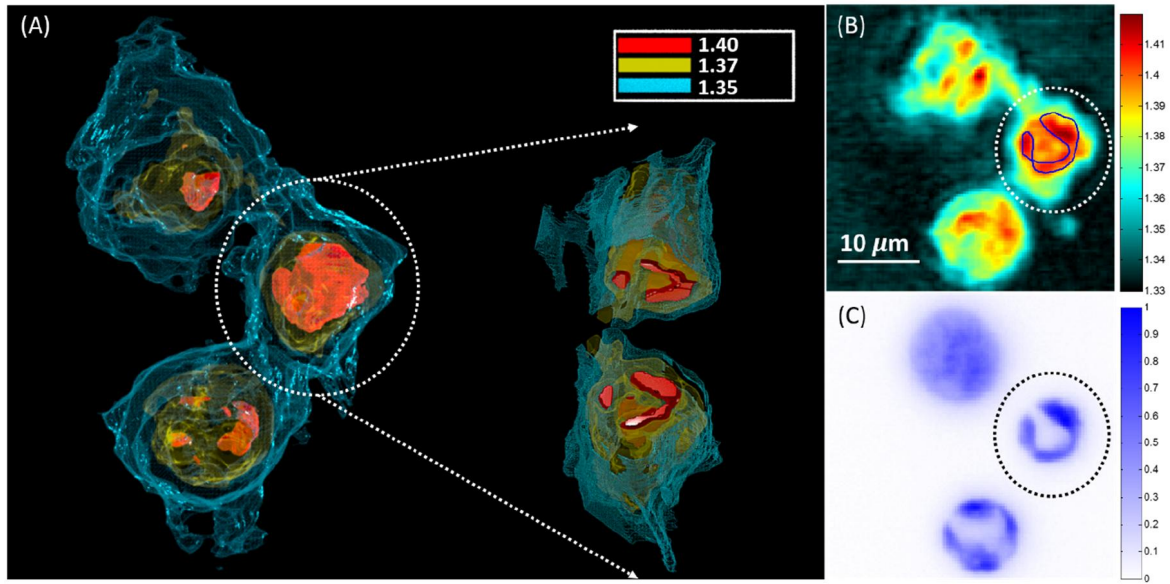

**Fig. 4. Comparison of apoptotic human Jurkat T cells between distributions of reconstructed RI values and wide-field fluorescence intensities. The Jurkat T cells are stained with Hoechst.** (A) Jurkat T cells with internal structures delineated from groupings of refractive index values (top view). Insert gives an open up view of the cell through the central z plane of the nucleus. (B) 2D slice of RI distributions alone z-axis. (C) Wide-field fluorescence intensity distributions of the same cells. The position of nuclear contents is highlighted with blue contour lines in (B) based on the intensity distributions of fluorescence (C).

**Supplementary Materials:**

Materials and Methods

Supplementary Text

Figures S1-S6

Movies S1-S3

# Supplementary Materials for

## Label-free three-dimensional (3D) structural imaging in live cells using intrinsic optical refractive index

Chen Liu[1,2,6], Michael Malek[2], Ivan K. H. Poon[3], Lanzhou Jiang[3], Colin J. R. Sheppard[4], Ann Roberts[5], Harry Quiney[5], Douguo Zhang[6], Xiaocong Yuan[1*], Jiao Lin[1,5,7], Christian Depeursinge[8,9*], Pierre Marquet[10,11,12*], Shanshan Kou[2,5,8*]

*Correspondence to: xcyuan@szu.edu.cn, christian.depeursinge@epfl.ch, Pierre.Marquet@neuro.ulaval.ca & s.kou@latrobe.edu.au

**This PDF file includes:**

Materials and Methods
Supplementary Text
Figs. S1 to S6
Captions for Movies S1 to S3

**Other Supplementary Materials for this manuscript includes the following:**

Movies S1 to S3

## Materials and Methods

### Culturing protocols

Primary cultures of cortical neurons were prepared from E17 OF1 mice embryos. Briefly, embryos were decapitated and brains removed and placed in PBS-glucose. Cortices were removed under a dissecting microscope and collected in a small Petri dish in PBS-glucose. Single-cell suspension was obtained by gentle trituration with a fire-polished Pasteur pipette in Neurobasal medium supplemented with B27 and GlutaMAX (Invitrogen). Cells were plated at an average density of 15000 cells/cm2 in supplemented Neurobasal medium on poly-ornithine coated glass coverslips (20mmϕ). After 3–4h, coverslips were transferred to dishes containing glial cell monolayers in supplemented Neurobasal medium. Neurons were maintained at 37°C in a humidified atmosphere of 95% air /5% $CO_2$ and were used after 21-35 days in vitro (DIV).

### Reagents

GSK 269962 (4009) was purchased from Tocris bioscience, UK. Trovafloxacin (PZ0015) and Hoechst 33342 (B2261) were purchased from Sigma-Aldrich, MO.

### Mammalian cell culture

Jurkat T cell lines were obtained from ATCC and cultured in complete RPMI media. Culture media were prepared using RPMI 1640 medium (Life Technologies, 22400-089), supplemented with 10% (vol/vol) fetal bovine serum (Gibco, 10099-141), penicillin (50U/ml) and streptomycin (50 mg/ml) mixture (Life Technologies, 15140122) and 0.2% (vol/vol) MycoZap (Lonza, Switzerland). Jurkat T cells were cultured at 37°C in a humidified atmosphere with 5% $CO_2$ at all times.

### Induction of apoptosis

Jurkat T cells were resuspended to about $3 \times 10^5$ cells/ml in 1% BSA/RPMI 1640 media and incubated in culture plates. Apoptosis was induced by UV irradiation with a Stratagene UV Stratalinker 1800 (Alient Technologies) at 150 mJ/cm$^2$ and incubated at 37°C in a humidified atmosphere with 5% $CO_2$ for about 2 h.

### Induction of secondary necrosis

After induction of apoptosis by UV irradiation, Jurkat T cells were cultured in 1% BSA/RPMI 1640 for 8 h. At this time point, the majority of apoptotic cells have progressed to secondary necrosis.

### Staining of the nucleus

Jurkat T cells were stained with Hoechst 33342 according to manufacturer's instructions to monitor the location of nuclear contents.

### Modulating the cell disassembly process during apoptosis

During apoptosis, fragmentation of the nucleus is driven by ROCK1-mediated membrane blebbing (*14*). In this study, Jurkat T cells were treated with the ROCK1 inhibitor GSK269962 (1μM) during apoptosis to inhibit the fragmentation of the nucleus caused by membrane blebbing so that the condensation of the chromatin can be monitored (*15*). For some experiments, Jurkat T cells were treated with trovafloxacin (20μM) during apoptosis to block the PANX1 membrane channel to promote the formation of apoptotic bodies (*16*).

**Supplementary Text**

Derivations in the cellular tomography

The classical derivations are briefly summarized below for the convenience of introducing our modification of the theory. Starting from the monochromatic Helmholtz equation ignoring the dielectric constant (*9*)

$$\nabla^2 \mathbf{E}(\mathbf{r}) + k^2 n^2(\mathbf{r})\mathbf{E}(\mathbf{r}) = 0 \tag{S1}$$

considering scalar condition for a single Cartesian component of $\mathbf{E}(\mathbf{r})$ as $U(\mathbf{r})$, and re-write the above Eq. S1 using the *scattering potential* of the medium defined by $F(\mathbf{r}) = k^2[n^2(\mathbf{r}) - 1]/4\pi$, we have

$$\nabla^2 U(\mathbf{r}) + k^2 U(\mathbf{r}) = -4\pi F(\mathbf{r})U(\mathbf{r}). \tag{S2}$$

Expressing the three-dimensional total field $U(\mathbf{r})$ as the sum of incident field, $U^i(\mathbf{r})$ and of the scattered field $U^s(\mathbf{r})$,

$$U(\mathbf{r}) = U^i(\mathbf{r}) + U^s(\mathbf{r}) \tag{S3}$$

and using the fact that incident field satisfies the Helmholtz equation, we have a Green's function based solution of the scattered field $U^s(\mathbf{r})$

$$U^s(\mathbf{r}) = \int F(\mathbf{r}')U(\mathbf{r}')G(\mathbf{r} - \mathbf{r}')d^3\mathbf{r}' \tag{S4}$$

with the Green's function as the solution to the Helmholtz equation, i.e.

$$(\nabla^2 + k^2)G(\mathbf{r} - \mathbf{r}') = -4\pi\delta^3(\mathbf{r} - \mathbf{r}'), \tag{S5}$$

where $\delta^3(\mathbf{r} - \mathbf{r}')$ is the *three-dimensional Dirac delta function*. Eq. S4 is an integral equation within the scattering volume V. Wolf then chose the *outgoing free-space Green's function* as following, which is basically the spherical wave,

$$G(\mathbf{r} - \mathbf{r}') = \frac{e^{ik|r-r'|}}{|r-r'|}. \tag{S6}$$

Substitute the above into Eq. S4 then using a Weyl expansion of the spherical wave to give two half spaces, i.e. $z^+$ and $z^-$, it arrives the classical optical diffraction formula, with two half-spaces separated by $z^+$ and $z^-$ (Suppl. Fig. 1)

$$\tilde{F}(K_x, K_y, K_z^\pm) = \frac{f_z}{2\pi i}\tilde{U}^{(s)}(f_x, f_y; z^\pm; s_0)e^{\mp if_z z^\pm}, \tag{S7}$$

where $\tilde{F}$ denotes the Fourier transform of the *scattering potential*.

We differ by substituting another Green's function $G_{cell}$ as the Helmholtz operator,

$$G_{cell} = G^*(\mathbf{r} - \mathbf{r}') - G(\mathbf{r} - \mathbf{r}'), \tag{S8}$$

where $G(\mathbf{r} - \mathbf{r}')$ is defined the same as in Eq. S6, and $G^*(\mathbf{r} - \mathbf{r}')$ means the complex conjugate. Essentially, the Green's function $G_{cell}$ contains only the propagating waves (ignoring evanescent components) (*6, 7*), which is valid in most imaging conditions of the biological cells. It takes the shape of a spherical shell in the Fourier domain for monochromatic wavelength and satisfies the Rayleigh-Somerfield's radiation condition. For describing the image formation of single-shot holography that satisfies Bragg's diffraction conditions, particular case of optical transfer functions, i.e., coherent transfer functions (CTF) has been studied, and the resulting CTF is called the Ewald sphere (*17*). We notice that the in Eq. S8, $G_{cell}$ transforms exactly to a spherical cap after a 3D Fourier transform. Therefore, we have set-up the scattered field $U^s(\mathbf{r})$ with a specific Green's function and this unifies with the CTF theoretical framework that we have previously presented, and henceforth the Fourier synthesis of 3D diffraction can be approached using the differential geometry solutions (*11*).

Substituting Eq. S8 and Eq. S6 into Eq. S4, and making use of the *Born's 1st order approximations* for weak objects

$$U^s(\mathbf{r}) = -2iA^{(i)} \int F(\mathbf{r}')e^{ik\mathbf{s_0}\cdot\mathbf{r}'} \frac{\sin(k|\mathbf{r}-\mathbf{r}'|)}{|\mathbf{r}-\mathbf{r}'|} d^3\mathbf{r}'$$
$$= -2iA^{(i)}(f \otimes_3 G_{img})(\mathbf{r}) \quad (S9)$$

where $A^{(i)}$ is the incident amplitude, $\otimes_3$ is a 3D convolution operator, $f(\mathbf{r}) = F(\mathbf{r})e^{ik\mathbf{s_0}\cdot\mathbf{r}}$ and $G_{img}(\mathbf{r}) = \frac{\sin(k|\mathbf{r}|)}{|\mathbf{r}|}$. The integral interval of Eq. S9 is within the scattering volume V. Taking the 3D Fourier transform of both sides of Eq. S9 and making use of the Convolution Theorem,

$$\tilde{U}^s(\mathbf{K}) = -2iA^{(i)}\tilde{F}(\mathbf{K}-k\mathbf{s_0})\tilde{G}_{img}(\mathbf{K}) \quad (S10)$$

where $\tilde{F}(\mathbf{K})$ is 3D Fourier transform of the scattering potential $F(\mathbf{r})$, and $\tilde{G}_{img}(\mathbf{K})$ is a 3D delta function which corresponds to the Ewald sphere. Noticing in a background refractive index $n_m$ rather than free space, the scattering potential,

$$F(\mathbf{r}) = k^2[n^2(\mathbf{r}) - n_m^2]/4\pi$$
$$= k_m^2[n^2(\mathbf{r})/n_m^2 - 1]/4\pi \quad (S11)$$

where $k_m$ is the wave vector in the background medium, $k_m = n_m \frac{2\pi}{\lambda}$. The 3D refractive index distribution then can be calculated,

$$n(\mathbf{r}) \triangleq \left[n_m^2 + \frac{in_m^2}{\pi A^{(i)} k_m} \mathcal{F}_3^{-1}\{\tilde{U}^s{}_3(\mathbf{K})\delta_3(\mathbf{K}-k_m)\}\right]^{1/2} \quad (S12)$$

where $\tilde{U}^s{}_3(\mathbf{K})$ is 3D Fourier transform of scattered field, $\mathcal{F}_3^{-1}$ represents 3D inverse Fourier transform.

Simulation using numerically treated Born's diffraction integral

A numerical approach to calculate the 3D scattered field of a weak object can be devised using the Monte Carlo principles (*18*). Then the Born's diffraction integral (*9*) is rewritten as

$$U^s(\mathbf{r}) = \frac{1}{nf_v} \sum_{i=1}^n F(\mathbf{r}_i)e^{ik\mathbf{s_0}\cdot\mathbf{r}_i} \frac{e^{ik|\mathbf{r}-\mathbf{r}_i|}}{|\mathbf{r}-\mathbf{r}_i|} \quad (S13)$$

where $f_v = 1/V$ if $\mathbf{r} \in V$ and 0 if $\mathbf{r} \notin V$. $V$ is the scattering volume, $n$ is the number of random events in the numerical recipe. For higher orders 3D Born scattered fields,

$$U_m^s(\mathbf{r}) = \frac{1}{nf_v} \sum_{i=1}^n F(\mathbf{r}_i)U_{m-1}(\mathbf{r}_i) \frac{e^{ik|\mathbf{r}-\mathbf{r}_i|}}{|\mathbf{r}-\mathbf{r}_i|} \quad (S14)$$

where $f_v$ and *n* are under the same conditions of S13, *m* is the order number of 3D Born scattered fields.

The dimension of our testing object is 2.5μm in diameter at the top flat region and the depth is 2μm. We make the immersion medium with refractive index 1.3336 and object region 1.3736. Hence the maximum optical path length is around 0.8 radians under wavelength 633nm, and hence it satisfies Born's 1st order approximation condition.

For realistic experimental conditions, we have simulated the scattered field of this object under one-dimensional scanning illumination from 0° to 30° that corresponds to an objective with N.A. up to 0.5. Following our proposed methodology, the conjugated K-space is filled in from individual spherical caps and consequently the RI distribution is accurately retrieved even if only a 1D scan configuration is used.

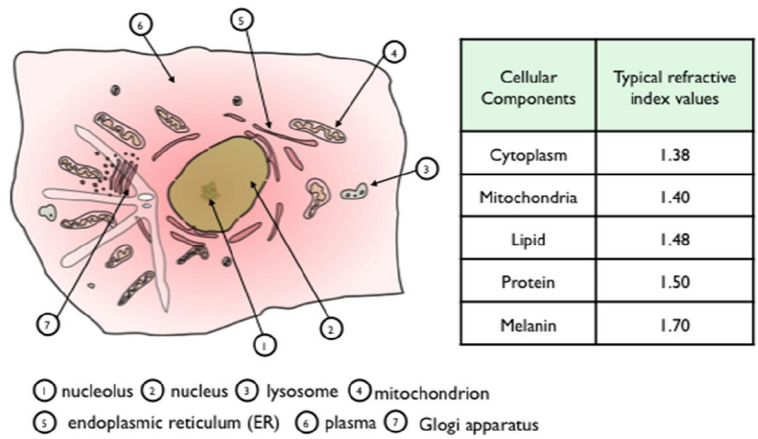

① nucleolus ② nucleus ③ lysosome ④ mitochondrion
⑤ endoplasmic reticulum (ER) ⑥ plasma ⑦ Glogi apparatus

**Fig. S1.**
Typical Eukaryote cellular structures and some known RI values of a few cellular constructs (*8, 19, 20*).

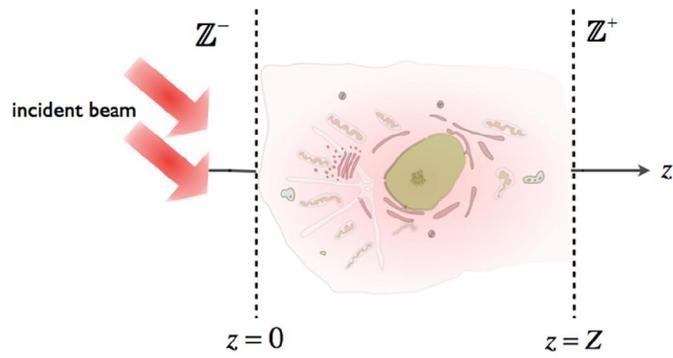

**Fig. S2**

Classical diffraction analysis of the three-dimensional semi-transparent object (*21*) breaks the spatial domain into two half spaces. The scattering potential is calculated based on known scattering distance $z^+$.

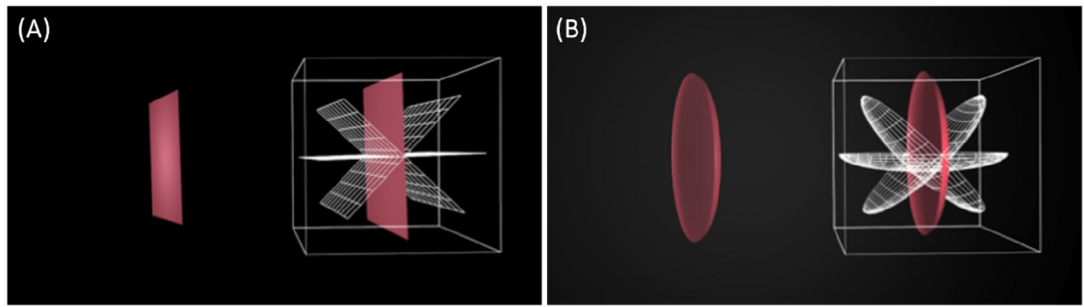

**Fig. S3**
The principle of x-ray computed tomography (CT) (*12*) (A) is fundamentally different from our cellular diffraction tomography (B) where the information in our case is represented on a spherical cap instead of a planar sheet.

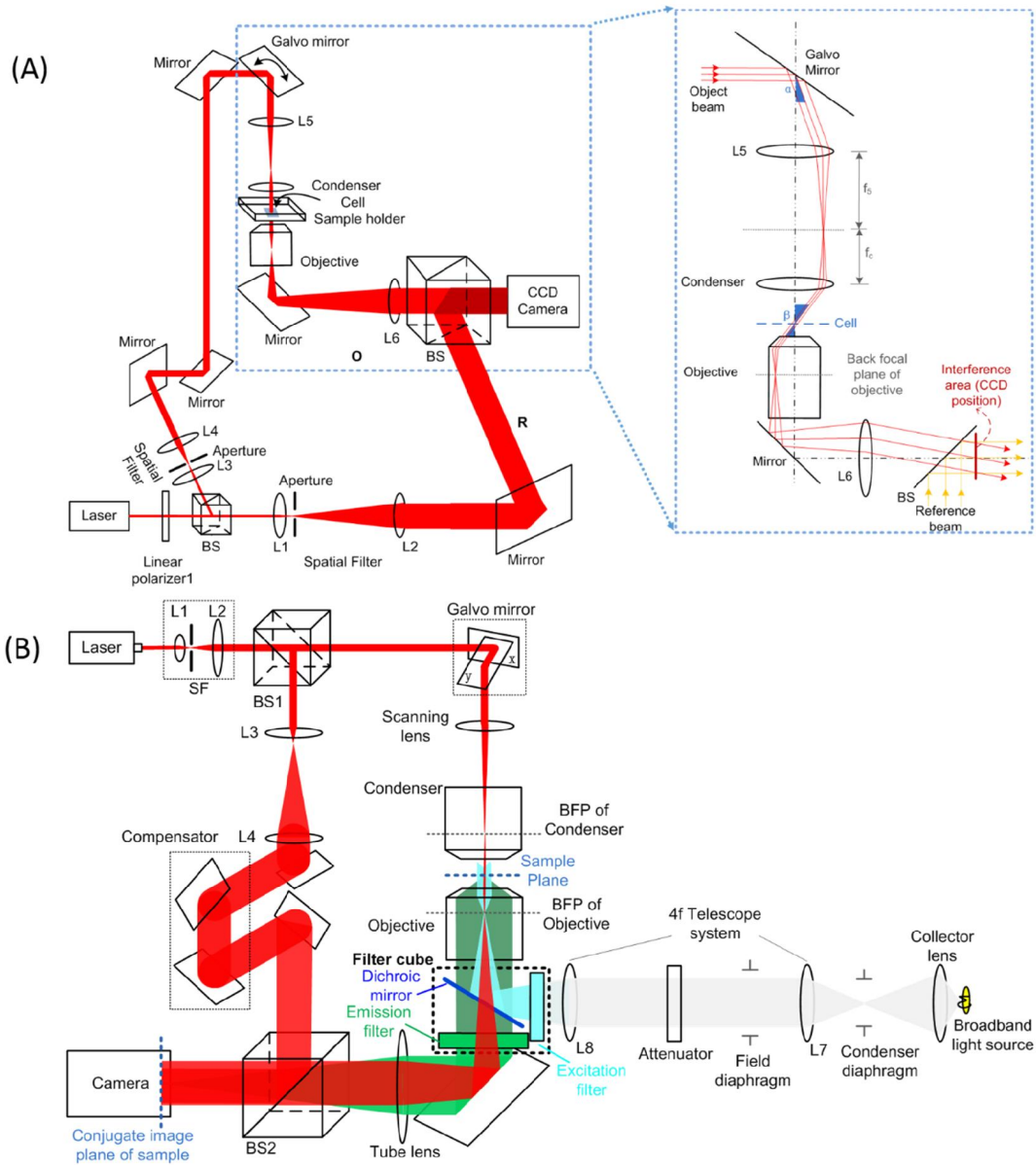

**Fig. S4**
(A) The cellular tomography is based on an off-axis digital holographic microscope setup in transmission. The collimated laser beam (at 635 nm) is divided by a beam splitter (BS) into two beams. The 3D microscopic biological sample is illuminated by one beam, and the microscope objective collects the transmitted light that forms the object wave (**O**). This object wave interferes at an angle (off-axis configuration) with the reference wave (**R**) to produce an image hologram intensity that is recorded by a high-speed CCD camera. Fast beam scans are enabled by optoelectronic galvo mirrors. L1, aperture and L2 forms spatial filtering. L3, aperture and L4 forms spatial filtering. L6 is an imaging lens. (B) Tomography setup combined with wide-field epi-fluorescence microscope. The magnification of objective is 40x (N.A. = 0.6).

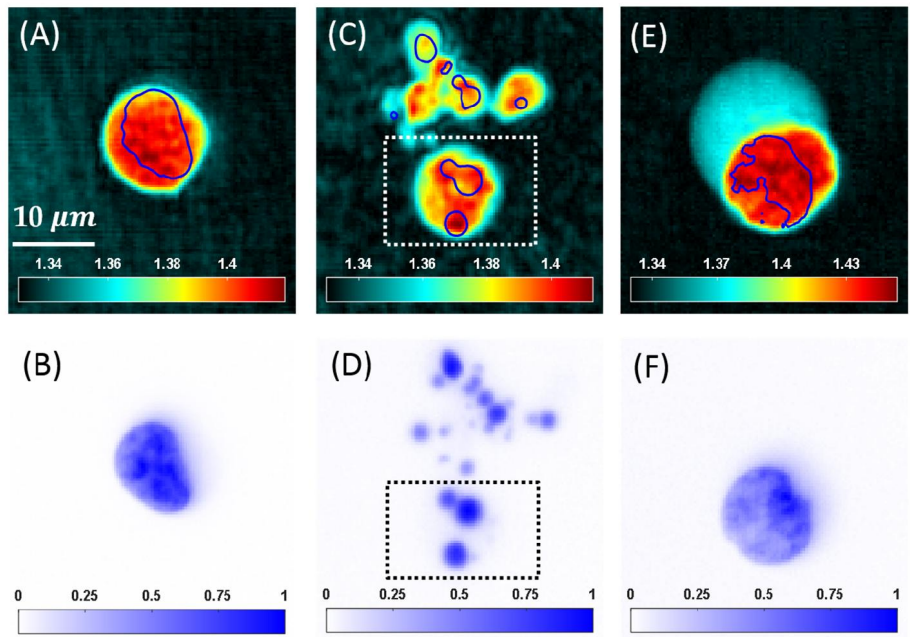

**Fig. S5**

Comparisons between the distributions of RI values and fluorescence intensities of Jurkat T cells under different conditions. The Jurkat T cells are stained with Hoechst so the fluorescence intensities here is a measure of nuclear contents. (A), (B) Viable Jurkat T cell; (C), (D) UV light-induced apoptotic Jurkat T cells with trovafloxacin treatment; (E), (F) Necrotic Jurkat T cell. The positions of stained DNA, which lots of proteins are localized with, are highlighted with blue contour lines in RI mappings based on each intensity distributions of fluorescence.

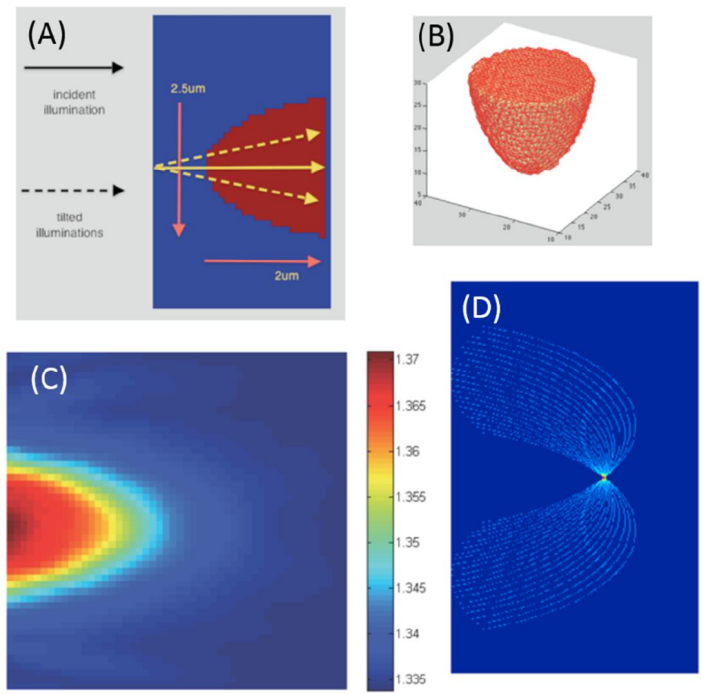

**Fig. S6**
Monte Carlo simulation of the phantom object volumetric diffraction and its quantitative reconstruction. (A) The 3D semi-spherical phantom object. (B) 3D view of the phantom object. (C) Reconstructed phantom object (side view) (D) Fourier "caps" synthesized with 21 angles.

**Movie S1**
The rotational view of the stain-free cheek cells with internal structures delineated from groupings of refractive index values.

**Movie S2**
The rotational view of the stain-free neuron with internal structures delineated from groupings of refractive index values.

**Movie S3**
The rotational view of ROCK inhibitor treated apoptotic Jurkat T cells with internal structures delineated from groupings of refractive index values.